\documentclass[aps,epsfig,eqsecnum]{revtex4}
\newcommand\be{\begin{eqnarray}}
\newcommand\ee{\end{eqnarray}}
\newcommand\ba{\begin{array}}
\newcommand\ea{\end{array}}

\def\r{\rangle}
\def\l{\langle}
\def\T{{\rm Tr}}

\def\cH{{\cal H}}

\def\cC{{\cal C}}
\def\cI{{\cal I}}

\def\cT{{\cal T}}

\def\cF{{\cal F}}

\def\cE{{\cal E}}

\def\cG{{\cal G}}

\def\cN{{\cal N}}
\def\cA{{\cal A}}

\begin{document}
%%%%%%%%%%%%%%%%%%%%%%%%%%%%%%%%%%%%%%%%%%%%%%%%%%%%%%%%%%%%%%%%%%%%%%
\title{Universality and optimality of programmable quantum processors}
\author{M\'ario Ziman$^{1,2,3}$ and Vladim\'\i r Bu\v zek$^{1,2,3}$}
\address{
$^{1}$~Research Center for Quantum Information, Slovak Academy of Sciences, D\'ubravsk\'a cesta 9, 845 11 Bratislava, Slovakia\\
$^{2}$~{\em Quniverse}, L{\'\i}\v{s}\v{c}ie \'{u}dolie 116, 841 04 Bratislava, Slovakia\\
$^{3}$~Faculty of Informatics, Masaryk University, Botanick\'a 68a, 602 00 Brno, Czech Republic\\
}
\begin{abstract}
We analyze and compare the optimality of approximate and probabilistic universal programmable quantum processors. We define several characteristics how to quantify the optimality and we study in detail performance of three types of programmable quantum processors based on (1) the C-NOT gate, (2) the SWAP operation, and (3) the model of the quantum information distributor - the QID processor. We show under which conditions the measurement assisted QID processor is optimal. We also investigate optimality of the so-called U-processors and we also compare the optimal approximative implementation of U(1) qubit rotations with the known probabilistic implementation as introduced by Vidal, Masanes and Cirac [ {\em Phys. Rev. Lett.} {\bf 88}, 047905 (2002)].
\end{abstract}

%\keyword{quantum computation, quantum programmability}


%%%%%%%%%%%%%%%%%%%%%%%%%%%%%%%%%%%%%%%%%%%%%%% BEGIN TITLE %%%%%%%%%%%%%
\maketitle
%%%%%%%%%%%%%%%%%%%%%%%%%%%%%%%%%%%%%%%%%%%%%%% END TITLE %%%%%%%%%%%%%%%%%%

%%%%%%%%%%%%%%%%%%%%%%%%%%%%%%%%%%%%%%%%%%%%%%%%%%%%%%%%%%%%%%%%%%%%%%%%%%%%%
\section{Programmable processors}
%%%%%%%%%%%%%%%%%%%%%%%%%%%%%%%%%%%%%%%%%%%%%%%%%%%%%%%%%%%%%%%%%%%%%%%%%%%%%
Classical programmable processors are realized as a hardware that
perform an operation (called computation) on a data register
according to instructions (program) encoded in a program register (software).
It is one of the central issues of computer science whether there
exist a universal (classical) processor that performs all possible
classical transformations of the data register of the size of $d$
bits \cite{gruska}. Let us, for simplicity, consider only reversible
classical computation (though
the conclusions are valid also for irreversible classical programs).
The register composed of $d$ classical bits can be reversibly transformed
in $2^d!$ different ways (permutations), e.g. we consider that
on a single bit we apply only two programs: the identity and the NOT operation.
For a single-bit data register the controlled-NOT (CNOT) gate serves
as a universal classical processor [the CNOT gate
is defined as follows ${\rm CNOT}: j,k\to j,j\oplus k$ ($j,k=0,1$)].
That is, if the program register
(consisting of a single bit) has the value (is in the state)
$k=0$ then the identity operation is realized on the data bit. Similarly, if $k=1$ then the NOT operation
is performed on the data bit. It is clear that such "control" devices realizing different
programs for different bit values of the program register, are universal
programmable processors also for larger data registers. The size of the
program register (in terms of the number $N$ of bits) is given by the relation
$2^d! = 2^N$, because $2^N$ represents the number of different
states of the register of the size $N$. As a result
we obtain that universal processor for $d$-bit data register
consists of approximately $N\approx 2^d(d-1)$ bits, i.e. it
is exponentially large.

In their seminal paper \cite{niel_chuang} Nielsen and Chuang showed
that the {\em quantum} analogue of universal programmable processor
does not exist, i.e. it is not possible even in principle to design
a universal deterministic programmable quantum processor. Information
about the quantum operation cannot be encoded into
the state of arbitrarily large program register. In order to
realize $n$ unitary transformation of the data register
one must use $n$ dimensional program register. The same holds
also in the classical case, but unlike there, in the quantum case even
for a single qubit the number of possible programs
(unitary transformations) is uncountably infinite. This requires
inseparable quantum systems as program registers, but such systems
are usually excluded from the standard quantum theory. The nonexistence
of universal programmable quantum device is another example of
no-go theorems in quantum information processing.

Let us consider two completely positive maps $\cE,\cF$ given by Kraus operators
$\{E_j\},\{F_k\}$, respectively.
In addition, let us assume that these two operations can be realized with a fixed processor
$G$, i.e. a unitary operation acting on the joint data plus program Hilbert
space $\cH_d\otimes\cH_p$. Denote $|\Xi_E\r,|\Xi_F\r$ the corresponding
pure states of the program register, i.e.
$\cE[\varrho]= \sum_j E_j\varrho E_j^\dagger \ \ \ {\rm with}\ \ E_j = \l j|G|\Xi_E\r$, where
$\cF[\varrho]= \sum_j F_j\varrho F_j^\dagger \ \ \ {\rm with}\ \ F_j = \l j|G|\Xi_F\r$, and $\{|j\r\}$ is some fixed basis of $\cH_p$.
Calculating $\sum_j E_j^\dagger F_j$ one derives the following identity
\be
\label{compatibility}
\sum_j E_j^\dagger F_j = \l\Xi_E|\Xi_F\r I = c I \, .
\ee
This equation is necessary for simultaneous realization of both
operations $\cE,\cF$ on the processor $G$. We remind us the
ambiguity of Kraus decomposition, i.e. when applying this criterion
one must take into account all possible decompositions. A special
case is achieved when $c=0$, i.e. the encoding program states are
orthogonal. In such a way any finite number of quantum
operations can be realized by encoding them  into mutually orthogonal states.
Let us apply this condition to the case of unitary operations $U_1,\dots,U_n$.
For them the Kraus decomposition is unique and we obtain the set of
conditions $U_j^\dagger U_k = c_{jk} I$. Obviously $c_{jj}=1$, but
in all other cases ($j\ne k$) it is necessary to set $c_{jk}=0$ in order
to satisfy Eq.~(\ref{compatibility}). It means that encoding $n$ unitaries requires
$n$-dimensional program space. By construction it can be seen that
such dimension of the program Hilbert space is also sufficient. One can simply define the processor
as $G=\sum_j U_j\otimes |j\r\l j|$. Using this criterion of compatibility
(\ref{compatibility}) one can investigate the programmability of
different families of processes. For instance, in Ref.~\cite{hillery2002b}
we have shown that
the family of phase damping channels can be implemented on a quantum processor,
but the family of amplitude damping channels cannot.

Properties of quantum programmable processors have been studied
already by many authors and from different perspectives
\cite{vlasov,dusek,dariano,geabanacloche1,deutsch,ekert,paz}.
In the present paper we focus our attention on optimality and universality of
approximate and probabilistic programmable quantum processors.
 The paper is organized as follows: In the Section II we
study optimality of programmable quantum processor, in the Section III we
analyze in detail optimality of three models of programmable quantum processors: the CNOT, the QID and
the SWAP processors. The Section IV is devoted to approximative programming. We will
show in which sense the QID processors are optimal.
In the Section V we relax the universality condition and we discuss programmable processor that allows us to implement
one-parametric group of single-qubit rotations.
We conclude our paper in Section VI with a brief summary of our results.

%%%%%%%%%%%%%%%%%%%%%%%%%%%%%%%%%%%%%%%%%%%%%%%%%%%%%%%%%%%%%%%%%%
\section{Universality and optimality}
%%%%%%%%%%%%%%%%%%%%%%%%%%%%%%%%%%%%%%%%%%%%%%%%%%%%%%%%%%%%%%%%%%

Even though universal deterministic programmable quantum processors do not exist \cite{niel_chuang} one can
investigate various approximations of these processors (this is a general approach when one deals with quantum-mechanical
no-go theorems).
One can study scenarios how to achieve
the universality by relaxing few of the ideal
conditions. In principle, there are two options:
i) an approximative implementation of quantum programs, or ii) a probabilistic
implementation of quantum programs. In the first case we allow some imprecision
$\epsilon$ in the implementation of the desired quantum programs, whereas
in the second case we relax the condition that the programmability
is deterministic by introducing a concept of the success probability
$P_{\rm success}$. Important point is that even though the implementation of the program is only
probabilistic, the measurement outcomes tell us exactly
when the desired operation is performed. It is not difficult to
see that in both of these cases the {\em universal} programmable processors
(either approximate or probabilistic) do exist
\cite{vlasov,dariano,hillery2002a,hillery2004,hillery2006}. However, since
the universality is conditioned by some imperfections the question
of optimality of encoding of quantum operations into states of quantum program registers is of importance
\cite{dariano,hillery2006,dariano2005}.

For a given processor $G$ one can should study whether it
is universal in an approximate, or in a probabilistic sense (or
both). Different (approximate or probabilistic) universal programmable
processors can be compared with the help of the approximation parameter
$\epsilon$, or the probability success $P_{\rm success}$, respectively.
There are several ways how to characterize the optimality
via these parameters. Let remind us the exact definition
of these parameters:

The approximate programmable processors perform a quantum
operation $\cT$ with the precision
$\epsilon(\cT)=\min_\xi D(\cT,\cE_\xi)$, where
$\cE_\xi [\varrho]= \T_p G(\varrho\otimes\xi)G^\dagger$ and
$D(.,.)$ quantifies the distance between two quantum operations.
In the probabilistic case we perform a measurement on the program
register at the output of the processor. In principle, we can distinguish two cases:
either the measurement is fixed (measurement-assisted processor),
or the choice of an arbitrary von Neumann measurement is a part
of the quantum programming. After recording an outcome $m$ of the measurement of the program register
the data register is transformed into the state
$\varrho_m=\cT_m[\varrho]=\frac{1}{p_m}\cI_m[\varrho]$, where
$p_m=\T\cI_m[\varrho]$ and $\cI_m$ is a linear completely
positive, but not necessarily trace-preserving, map. Only in cases
when $p_m\ne p_m(\varrho)$ the transformation
$\cT_m$ corresponds to some quantum operation, i.e. it is a
completely positive trace-preserving linear map. Without the measurement,
or better to say without the post-selection, the data register is transformed
by some quantum operation $\cE_\xi$. These maps are always performed
with a probability equal to unity. For each program state one can express
the realized operation as a convex combination
$\cE_\xi=\sum_j q_j \cT_j$ and the problem is
whether it is possible to find a measurement $M$
such that the operation $\cT_j$ was realized with the probability
$q_j$. The decomposition of $\cE_\xi=\sum_m q_m\cT_m$
is called realizable if there exists a measurement $M$
of the program register with outcomes $m$ such
that $\cT_m=\frac{1}{p_m}\cI_m$.
The success probability $P_{\rm success}(\cT)$ of the
operation $\cT$ is defined as the maximum of probabilities $p$
over all program states $\xi$ with the realizable decomposition
$\cE_\xi=p\cT+(1-p)\cN$. The situation is simple if one
uses the measurement-assisted quantum processor, i.e. the measurement
is fixed for all inputs. The universality of such device was demonstrated
explicitly in Ref. \cite{hillery2002a}, but the optimality is still an open
question. It is clear that limits of measurement-assisted
processors are stronger than limits for
probabilistic processors where one can vary measurements in order
to increase success probabilities.

There are several approaches how to compare performance
of quantum programmable processors. One can use either extremal
(worst) cases, or average values to evaluate the accuracy/success of
approximate/probabilistic processors, i.e.
\be
\nonumber
\overline{P}_{\rm success}^G=\int_\cT {\rm d}\cT P_{\rm success}(\cT),\ & \
\overline{\epsilon}_G=\int_\cT {\rm d}\cT \epsilon(\cT)\\
\nonumber
P_{\rm success}^G = \min_{\cT} P_{\rm success}(\cT), \ & \
\epsilon_G=\max_\cT \epsilon(\cT)
\ee
There is also a freedom
in the choice of the function $D(.,.)$ in the definition of the accuracy $\epsilon(\cT)$ of
the approximation. The larger the
success probability the better the probabilistic processor and similarly, the
smaller the error the better the approximate processor.
We will use the notation $P_{\rm success}^{G,M}$ and
$\overline{P}_{\rm success}^{G,M}$ for the parameters of the
measurement-assisted quantum processor specified by the unitary
transformation $G$ and the program-register measurement $M$.
The following relations hold:
$\max_M P_{\rm success}^{G,M}\le P_{\rm success}^G$ and
$\max_M \overline{P}_{\rm success}^{G,M}\le \overline{P}_{\rm success}^G$.

Also the universality of quantum processors is usually understood in two
different ways: i) either with respect to an implementation of all
quantum operations, or ii) with respect to a realization of all
unitary transformations of the program register.
In some cases the set of implemented operations
can be reduced (restricted) to smaller families of quantum
operations. It is clear that due to a unitary representation
of any completely positive tracepreserving linear map both
of the meaning of universality are closely related. In order to
define optimal quantum processor we usually fix the size of the
data and program register, $d={\rm dim}\cH_d$ and $N={\rm dim}\cH_p$.
The main open problem of quantum processor's optimality
is to find this functional dependence for success probabilities
and approximation accuracy. In other words, for a fixed length
of data and program register the task is to find the class of processors
maximizing the success probability and minimizing the approximation
parameter.

Let us take a process fidelity to quantify the distance
between two maps, i.e. $D(\cE,\cT)=1-F(\cE,\cT)=1-f(\Phi_\cE,\Phi_\cT)$,
where $\Phi_\cE=\cE\otimes\cI[\Psi_+]$ ($\Psi_+$ is the projector
onto maximally entangled state $|\psi_+\r
=\frac{1}{\sqrt{d}}\sum_j |j\r\otimes|j\r$) and
$f(\varrho,\sigma)=(\T\sqrt{\sqrt{\varrho}\,\sigma\sqrt{\varrho}})^2$ is the
state fidelity function. From the definition of the
quantities $P_{\rm success}(\cT)$, $\epsilon(\cT)$
and properties of process fidelity it follows that \cite{hillery2006}
\be
P_{\rm error}(\cT)\ge\epsilon(\cT)\, ,
\ee
where $P_{\rm error}(\cT)=1-P_{\rm success}(\cT)$. The probabilistic
realization of $\cT$ means that a given program state $\xi$ induces
the map $\cE_\xi=P_{\rm success}\cT+P_{\rm error}\cN$. One can
say that program encoded in state $\xi$ approximates
the transformation $\cT$ with the
precision quantified by $\epsilon=1-F(\cE_\xi,\cT)$. Using the concavity of
process fidelity one can directly show that
$\epsilon\le 1-P_{\rm success}+P_{\rm error} F(\cN,\cT)\le 1-P_{\rm success}$
which gives the above inequality.

Similar relations hold also for the derived average, and worst case
quantities. This inequality is not saturated in general (for fixed
quantum processor), however it might be the case that for the
optimal values (optimized over all processors)
these two numbers coincide. It could be an interesting
result if, moreover, the same optimal processors are optimal
for approximate as well as probabilistic scenario.

For our purposes it will be useful to have an expression
for the process fidelity
between an arbitrary channel $\cE$ (defined as
$\cE[\varrho]=\sum_r A_r\varrho A_r^\dagger$) and a unitary transformation
$U$. The state $\Phi_U=U\otimes I[\Psi_+]$ is pure, i.e.
$\sqrt{\Phi_U}=\Phi_U$. It follows that
\be
\nonumber F(U,\cE)=\l\Phi_U|\cE\otimes\cI[\Psi_+]|\Phi_E\r
=\frac{1}{d^2}\sum_{j,k}\l j|U^\dagger\cE[|j\r\l k|]U|k\r
=\frac{1}{d^2}\sum_r |\T U^\dagger A_r|^2 \, .
\ee

\section{Programmability of unitary transformations}

In this section we will pay attention to a simpler problem
of implementation of all unitary maps. In such case there exists a unique
Haar measure $dU$ that enables us, in principle, to calculate
also the average error and average success probability. In what follows
we shall analyze three examples of quantum processors.

\subsection{Controlled NOT}
At the beginning of this paper we have seen that the controlled NOT (CNOT)
gate serves as universal classical programmable processor implementing the
programs on a single classical bit. Its quantum version,
$G_{\rm CNOT}=I\otimes |+\r\l +|+\sigma_z\otimes|-\r\l -|$
(with $|\pm\r=\frac{1}{\sqrt{2}}(|0\r\pm|1\r)$),
is not universal, but still it can serve as a good
and simple example of a quantum processor realizing approximatively
and probabilistically a specific subclass
of unitary operations. Using this device we are able to realize
deterministically arbitrary channel of the form
$\cE[\varrho]=p\varrho+(1-p)\sigma_z\varrho\sigma_z$.
Vidal and Cirac \cite{vidal1,vidal2} showed that
using the measurement-assisted CNOT processor
one can probabilistically implement arbitrary
$U(1)$ rotation $U_\varphi=\exp(-i\varphi\sigma_z)$ with the probability
$P_{\rm success}(U_\varphi)=1/2$. The fixed measurement is
specified by the basis $|0\r,|1\r$.

In Ref.~\cite{ziman2003} we studied different choices of measurement
for the CNOT processor. We found that arbitrary measurement
enables us to realize probabilistically a whole set of
unitary transformations $U_\varphi$, though in this case probability distributions
$P_{\rm success}^{M}(U_\varphi)$ are different for different measurements $M$.
Of course, the transformations $I,\sigma_z$ are always implemented with
the probability equal to. The success probability for a fixed measurement $M$
is given by the relation $P_{success}=\cos^2\xi\cos^2\eta+\sin^2\xi\sin^2\eta$,
where the angle $\eta$ specifies the choice of the measurement
and $\xi$ represents the states encoding the unitary transformations
$U_\varphi$ for $\varphi=\arccos(\frac{\cos\xi\cos\eta}
{\sqrt{\cos^2\xi\cos^2\eta+\sin^2\xi\sin^2\eta}})$. Except the cases
when $\eta=k\pi/2$ (for $k=0,1,2,3,\dots$), each unitary
operation $U_\varphi$ is realized. Therefore,
either $I$, or $\sigma_z$ is realized on measurement-assisted CNOT processor
with the smallest probability depending
on whether $\sin^2\eta$ is larger than $\cos^2\eta$, or not. In particular
$P_{\rm success}^{{\rm CNOT},\eta}=\min\{\cos^2\eta,\sin^2\eta\}$.
The average success probabilities for measurement-assisted
CNOT are given as follows
\be
\nonumber
\overline{P}_{\rm success}^{{\rm CNOT},\eta}
=\frac{1}{2\pi}\left(\int_0^{2\pi} {\rm d}\xi\sin^2\xi
+\cos^2\eta\int_0^{2\pi}{\rm d}\xi \cos 2\xi\right)
=\frac{1}{2}\, .
\ee
A straightforward calculation shows that in the {\em average} sense
all measurement-assisted CNOT processors are equivalent. On the other hand  they
are not equivalent in the sense of the worst success probabilities.

It is not easy to calculate these quantities for the case when we are
allowed to alternate (optimize) measurements in order to increase the success
probability. In particular, it is difficult to express analytically
the dependence $\xi=\xi(\varphi)$ and consequently to find an analytic
expression for the optimal success probability $P_{\rm success}(U_\varphi)$.
It is clear that each of the transformations can be realized
with the probability strictly larger than $1/2$, i.e.
$P_{\rm success}^{\rm CNOT}>1/2$ and also
$\overline{P}_{\rm success}^{\rm CNOT}>1/2$.

Let us assume that we want to perform unitary transformations
$U_\varphi$ approximatively. What is then the optimal program state $\xi_\varphi$
approximating the given $U_\varphi$ on the CNOT processor?
Denoting by $\cE_p[\varrho]=p\varrho+(1-p)\sigma_z\varrho\sigma_z$
and using the derived expression for the process fidelity between
an arbitrary channel and a unitary transformation we get
for the error
\be
\epsilon(U_\varphi)&=&1-\max_p F(U_\varphi,\cE_p)
=1-\frac{1}{4}\max \{|\T U_\varphi^\dagger|^2,
|\T U_\varphi^\dagger\sigma_z|^2\}\, ,
\ee
i.e. it is optimal to encode the operation $U_\varphi$
into one of two program states encoding the identity and $\sigma_z$
operation, i.e. state $|\pm\r$. The encoding is optimal for
the state for which the overlap between the desired transformation
and the program $I$ (or $\sigma_z$) is larger. In particular,
$\epsilon(U_\varphi)=1-\max\{\cos^2\varphi,\sin^2\varphi\}$.
That is, the characteristics of the approximate programmable
CNOT processor describing the quality of the realization of the
set of unitary transformations $U_\varphi$ are the following
\be
\epsilon_{\rm CNOT}&=& \max_\varphi\epsilon(U_\varphi)=1/2\, ,\\
\overline{\epsilon}_{\rm CNOT}&=&
\frac{1}{2\pi}\int {\rm d}\varphi\,\epsilon(U_\varphi)=1/2-1/\pi \, .
\ee
One can directly verify that the bound $P_{\rm error}\ge \epsilon$
holds in all its variations. It
is saturated only for the worst case  typical for the
optimal measurement-assisted CNOT processor compared with the
worst case approximation error, when we have
$P_{\rm error}^{{\rm CNOT},\eta=\pi/4}=\epsilon_{\rm CNOT}=1/2$.

\subsection{Quantum information distributor}
Originally, the quantum information distributor (QID)
was introduced as a device
performing the optimal quantum cloning and the optimal quantum NOT operation.
In \cite{hillery2002a} the authors showed that the QID
can be used as a universal
measurement-assisted processor realizing all unitary
transformations. The QID is a device with a data register representing
by a qudit ($d$ dimensional quantum system)
and program register represented by two qudits, i.e. $N=d^2$.
It belongs to the family of U processors \cite{hillery2002b},
or equivalently the controlled-U gates.
It is defined as
$G_{\rm QID}=\sum_k U_k\otimes |\theta_k\r\l\theta_k|$,
where $k=0,\dots,d^2-1$, $U_k\equiv U_{ab}=\sum_r e^{i2\pi ar/d}|r-b\r\l r|$
(for qubits sigma operators), $|\theta_k\r=U_k\otimes I|\theta_0\r$,
$|\theta_0\r=\frac{1}{\sqrt{d}}\sum_j |j\r\otimes|j\r$. In other words,
the unitary operators $U_k$ form an
orthogonal operator basis $\T U_j^\dagger U_k=d \delta_{jk} I$, and the
states $|\theta_k\r$ form the basis of two-qudit Hilbert space
composed of maximally entangled states. In particular, for qubit
$U_k=\sigma_k$ and $\{|\theta_k\r\}$ is the Bell basis. The
program states $|\theta_k\r$ encode unitary transformations $U_k$.
A general program state $\xi$ induces one of the generalized Pauli channels
$\cE_\xi[\varrho]=\sum_k p_k U_k\varrho U_k^\dagger$ with
$p_k=\l\theta_k|\xi|\theta_k\r$.

Using the measurement basis
$|m_j\r\equiv|m_{xy}\r
=|-x\r\otimes\frac{1}{\sqrt{d}}\sum_r e^{i2\pi y/d}|r-x\r$
one can probabilistically implement an arbitrary unitary
transformation with the probability $P_{\rm success}=1/d^2$
\cite{hillery2004}. The result
$|m_{00}\r = \frac{1}{d^2}\sum_k |\theta_k\r =
|0\r\otimes\frac{1}{\sqrt{d}}\sum_r |r\r$ indicates the successful
realization of the program $U$. In particular, using this measurement
basis the action of the processor can be written in the following form
\be
G_{\rm QID}|\psi\r\otimes|\Xi\r =
\frac{1}{d^2}\sum_j U_j A(\Xi) U_j^\dagger |\psi\r\otimes|m_j\r \, ,
\ee
where $A(\Xi)=\sum_k \l\theta_k|\Xi\r U_k$. If $A(\Xi)$ is a unitary operation
then probabilities of outcomes $m_j$ are data independent and they read
$p(m_j)=\frac{1}{d^2}$. The described measurement-assisted quantum
processor implements unitaries with the success probability
\be
\overline{P}_{\rm success}^{{\rm QID},M}=P_{\rm success}^{{\rm QID},M}
=1/d^2\, .
\ee
It is of interest to show whether this is the optimal measurement-assisted
probabilistic realization, or not.
Note that this processor is indeed very specific, because it performs
all unitaries with the same probability $P_{\rm success}(U)=1/d^2$.
A similar conclusions like in the case of CNOT processor can be made.
In particular, this measurement-assisted QID processor is optimal in the
sense of worst case optimality, because using different
measurements will result in nontrivial success probability distributions
over the set of unitary transformations.

The approximate realization is again similar to the case of
CNOT gate. Both of them belong to the family of U processors, i.e.
they are of the form $G=\sum_j U_j\otimes |j\r\l j|$, where
$U_j$ are arbitrary unitary operations and program states
$\{|j\r\}$ form an orthonormal basis of $\cH_p$. These processors
are not very ``rich'' from the point of view of approximative implementation
of all unitary transformations. In fact, these processors
deterministically perform the random unitary channels
$\cE_{\vec{p}}[\varrho]=\sum_j p_j U_j\varrho U_j^\dagger$ with
$p_j=\l j|\xi|j\r$ providing that the program register was initially
prepared in the state $\xi$. As we shall see
for the approximative programming only the basis states $|j\r$
are useful. In fact
\be
\epsilon(U)
=\max_{\vec{p}} F(U,\cE_{\vec{p}})
=1-\frac{1}{d^2}\sum_j p_j|\T U^\dagger U_j|^2
=1-\frac{1}{d^2}\max_j\{|\T U^\dagger U_j|^2\}\, .
\ee
For a unitary $U$ the best approximation is achieved by
the program state $|j\r$ maximizing the overlap $|\T U^\dagger U_j|^2$.
For the QID processor an arbitrary $U$ can be written as a
linear combination of the operators $U_k$, i.e.
$U=\sum_k \alpha_k U_k$ and $\epsilon(U)
=1-\frac{1}{d^2}\max_j |\alpha_k \T U_k U_j|^2 =1-\max_j |\alpha_j|^2$.
Thus, the quality of approximations on
the QID processor is given by
\be
\epsilon_{\rm QID}=\max_U\epsilon(U)=1-1/d^2\, .
\ee
This value is achieved for the unitary transformation
$U=\frac{1}{2}\sum_j U_j$.

\subsection{SWAP}
The quantum SWAP gate acts on two registers of the same size,
$d=N$, and its performance is defined in the following way
$G_{\rm SWAP}(\varrho\otimes\xi)G^\dagger_{\rm SWAP}
=\xi\otimes\varrho$. Taking the SWAP gate as a processor we can implement
all contractions to a single point specified by the program state,
i.e. $\cC_{\xi}[\varrho]=\xi$ for all input states $\varrho$.
It is clear that such processor does not belong to the family of
U processors. Indeed it cannot be used to realize any unitary
transformation. However, unlike for U processors, for each program state
the SWAP processor implements different program, which
makes it exceptional, because there is no redundancy
in the state space of the program register. Let remind us that for
U processors all states having the same diagonal elements in the
basis of vectors $|j\r$ encode the same quantum operation.
The set of contractions is closed under convex combinations
and except extremal points of the whole set of quantum operations
(pure-state contractions) they contain also totally random channel,
i.e. a contraction to the total mixture.

Let us consider an arbitrary measurement $M$ of the program register.
The probability of measuring the result $|m\r$ is given as
\be
\nonumber
p(m)=\T[Q_m G_{\rm SWAP}(\varrho\otimes\xi)G^\dagger_{\rm SWAP}Q_m]
=\T [\xi \otimes |m\r\l m|\varrho|m\r\l m|]
=\l m|\varrho|m\r\, ,
\ee
where we used the notation $Q_m=I\otimes|m\r\l m|$. That is, for any
measurement the resulting probability distribution depends on
the initial state of the data, i.e. no specific quantum operation
is associated with particular results. Or, to be more precise, for
each outcome the contraction to a fixed point $\xi$ is realized. Therefore,
measurement-assisted SWAP processors are in some sense trivial and do not
provide us with any improvement of performance. Therefore , it does not make much sense
to apply measurements at the output of the SWAP processor. Consequently, the SWAP processor
is not suitable for performing probabilistically any unitary transformation, i.e.
$P_{\rm success}^{\rm SWAP}=\overline{P}_{\rm success}^{\rm SWAP}=0$.

In what follows we will see that from the point of view of approximative
implementation of unitary transformations, the SWAP processor acts
much better. Let us calculate the error
$\epsilon(U)=1-\frac{1}{d^2}\max_\Xi\sum_r |\T U^\dagger A_r(\Xi)|^2$,
where $A_r(\Xi)$ are Kraus operators associated with pure program
states, i.e. $A_r(\Xi)=\l r|G_{\rm SWAP}|\Xi\r
=\l r|\sum_{j,k}|k\r\l j|\otimes|j\r\l k|\Xi\r=|\Xi\r\l r|$.
In the matrix (basis) representation of $G_{\rm SWAP}$ one can use any basis, i.e.
even a basis containing the state vector $|\Xi\r$. Our task is to find
$\epsilon(U)=1-\frac{1}{d^2}\max_\Xi \sum_r |\l r|U^\dagger|\Xi\r|^2$.
Choosing the basis containing the vector $|\Xi\rangle$ we obtain that
we have to maximize the length of the column of the
unitary transformation $U^\dagger$. But we know that columns form
mutually orthonormal vectors, and therefore in this basis the
expression $\sum_r |\l r|U^\dagger|\Xi\r|^2=1$. This means that each pure
program state $|\Xi\r$ approximates each unitary operation
with the same accuracy measured by $\epsilon(U)=1-1/d^2$. And consequently
\be
\epsilon_{\rm SWAP}=\overline{\epsilon}_{\rm SWAP}=1-1/d^2 \, .
\ee
Thus, the worst case accuracy of the QID and the SWAP processors are the same. Nevertheless, it should be noted that
for the QID the program space is twice as large as the program
space for the SWAP processor. Therefore, we can conclude that the SWAP
processor is more optimal (suitable) for approximate programming. However,
the situation is completely different for probabilistic
programming.

%%%%%%%%%%%%%%%%%%%%%%%%%%%%%%%%%%%%%%%%%%%%%%%%%%%%%%%%%%%%%%%%
\section{Optimality of approximate processors}
%%%%%%%%%%%%%%%%%%%%%%%%%%%%%%%%%%%%%%%%%%%%%%%%%%%%%%%%%%%%%%%%
Each quantum processor induces a mapping $\cG$ from the set of program
states into a subset of all quantum operations applied on the data register,
i.e. $\cG:\xi\mapsto\cE_\xi$. Let us denote by $\Gamma_\cG\subset\Gamma$
the subset of deterministically implementable quantum programs, where
$\Gamma$ stands for the set of all possible quantum operations.
Since $\Gamma_\cG$ is a linear image of the set of program states, it is
convex. The question of optimality then can be illustrated in the following
way. Denote by $\partial\Gamma$ the boundary with respect to some
topology. Then the worst case optimality parameter measures
the distance between the points of the sets $\Gamma_\cG$
and $\partial\Gamma$. Formally,
$\epsilon_G=\max_{\cT\in\Gamma_\cG}\min_{\cE\in\Gamma} D(\cE,\cT)$.

For the process fidelity we have
\be
\epsilon_G=1-\min_{\cT\in\Gamma_\cG}\max_{\cE\in\Gamma} F(\cE,\cT)\, .
\ee
Let us consider $\cE^\prime,\cT^\prime$ that optimize the accuracy.
Because of the fact that $F(\sum_k p_k \cE_k,\cT^\prime)\ge \sum_k p_k F(\cE_k,\cT^\prime)$
it follows that $\cE^\prime$ can be always chosen to be the extremal
quantum operation of the set $\Gamma$. Let us formulate the main problem:
given a data register of size $d$ and given the program register of size
$N$. What is the optimal approximate processor? This problem is
indeed difficult and only partial results are known.

Let us pay attention to an optimal realization of all unitary transformations.
In this case the approximate processor is {\em universal} if it performs an arbitrary
unitary transformation with nonzero accuracy, i.e. $0< \epsilon(U) \leq 1$. For
instance, the CNOT processor is not universal in this sense, but the SWAP processor
is. The necessary and sufficient condition for a processing being universal is the following:
 The process fidelity has to nonzero for all unitary transformations which
means that $F(U,\xi)=\frac{1}{d^2}\sum_r |\T U^\dagger A_r(\xi)|^2>0$.
The general processor can be written in the form
$G=\sum_{jk} A_{jk}\otimes |j\r\l k|$. Providing that $A_{jk}$ form
a complete operator basis the process fidelity is different from
zero for all unitary transformations. In particular, for the SWAP processor
the operators $A_{jk}=|k\r\l j|$ undoubtedly form a complete basis.

For the $d$-dimensional data register the processor
can be universal (implementing all unitaries) only
if the operators $A_{jk}$ form an operator basis, i.e.
they are independent and the total number of them is $d^2$.
This necessarily means that universal processors
for the $d$ dimensional program registers must
use at least $d$ dimensional program register.
It is easy to see that U processors with such program size
cannot be universal, because they do not contain sufficient
amount of independent operators, $A_{jk}=\delta_{jk} U_j$.
The existence of an approximate universal processor of this
size is guaranteed by an example of the SWAP processor.
The question is whether this processor is optimal, i.e.
whether $\epsilon_{\rm SWAP}=1-1/d^2$ attains indeed the minimal
value for a universal processor of the program size $N=d$.
D'Arianno and Perrinoti \cite{dariano,dariano2005} found that this is indeed
the case for a single qubit (although they considered a different
processor).

In some sense such an error can be achieved trivially.
The process fidelity will be nonzero for all quantum operations
providing that at least for one of the program states $\xi$
encodes an operation $\cT$ such that the state
$\Phi_\cT$ has the full rank, i.e.
$\T\sqrt{\sqrt{\Phi_\cE}\Phi_\cT\sqrt{\Phi_\cE}}>0$ for all $\cE$.
This condition is only sufficient, but not necessary for a processor
to be universal. Anyway, once we are discussing the problem in general,
this condition is sufficient to derive the the lower bound on the accuracy
of approximate processors. Consider that the processor
enables us to perform a contraction into the total mixture, i.e. let us consider
the map $\cA[\varrho]=\frac{1}{d}I$. Then for any $\cE$ the process
fidelity reads $F(\cA,\cE)=\frac{1}{d^2}[\T\sqrt{\Phi_\cE}]^2$, because
$\Phi_\cA=\frac{1}{d^2}I$. The minimum
is achieved for pure states $\Phi_\cE$ corresponding to unitary
transformations. That is, all other programs are approximated by this
operation better than unitaries and therefore
for such universal processor $\epsilon_G\ge 1-1/d^2$. The SWAP
as well as the QID processors can realize the transformation $\cA$ and
they both saturate this bound.

Let us restrict the original problem and ask the following
question: What is the optimal U processor? To be able to discuss
optimality processors they have to be universal. It means that
this question does not make any sense for program registers of
the size $N<d^2$. The QID is an example of an approximate processor
in the case when $N=d^2$. For the U processors the situation is simpler because
the process fidelity is given by the formula
$F(U,\cE)=\frac{1}{d^2}\sum_r |\T U^\dagger A_r|^2$. Let us fix the
dimension of the program space to be $N=d^2$, i.e. we work with $d^2$
unitaries $U_j$. The first question is which of the
U processors is optimal in implementing all unitary
transformations. We know that the best approximation is achieved
for the basis states $|j\r$ and
$\epsilon(U)=1-\frac{1}{d^2}\max_{j}\{|\T U^\dagger U_j|^2\}$.
That is, the worst case is represented by a unitary $U_x$ having the
smallest maximal overlap with ``elementary'' programs $U_j$. Our aim
is to show that this worst case is optimized for mutually
orthogonal operators $\sigma_j$, i.e.
$\T\sigma_j^\dagger\sigma_k=d\delta_{jk}$. Processors
having as elementary programs mutually orthogonal unitaries
are all equivalent. In particular, the approximation of the operator
$U_x=\frac{1}{d}\sum_j \sigma_j$ maximize the error to
$\epsilon(U_x)=1-1/d^2$. Each unitary transformation can be expressed
in a suitable orthogonal operator basis in such form, i.e. each
unitary operation can represent the worst case for some
processor.

Consider that we know the transformation $U_x$ for a processor
with nonorthogonal unitaries. Let us define a processor
using orthogonal operators $\sigma_j$ such that
$U_x=\frac{1}{d}\sum_j \sigma_j$. Let us express also the
elementary operators $U_j$ in this new operator
basis $U_j=\sum_a u_{ja}\sigma_a$. Calculating the overlap
we obtain the following bound
\be
|\T U_x U_k|^2 =
|\frac{1}{d}\sum_j u_{kj}|^2\le \sum_j |u_{kj}|^2\le 1\, ,
\ee
where we used the identities $\T\sigma_a^\dagger\sigma_b=d\delta_{ab}$ and
$\T U_k^\dagger U_k=d\sum_j |u_{kj}|^2=d$. Consequently, the error
is minimized for $|\T U_x\sigma_k|^2=1$ and therefore we can conclude
that the U processors with orthogonal elementary programs are indeed
optimal, i.e. the QID processor is an optimal approximate U-processor
for $N=d^2$ dimensional program register.

This optimality result for approximate processors
can be used directly for probabilistic processors by using
the inequality $P_{\rm error}\ge \epsilon$, or equivalently
$P_{\rm success}\le 1-\epsilon$. The problem is now
the following: Which of the U processors is optimal
in a probabilistic realization of unitary transformations?
Based on the answer to the similar question for approximative
programming we can say that the optimal success probability
is less or equal to $1-\epsilon=1/d^2$, i.e. $P_{\rm success}\le 1/d^2$.
However, we know that the measurement-assisted QID processor saturates
this bound, i.e. $P_{\rm success}^{{\rm QID},M}=1/d^2$, and
therefore the QID is an example of the optimal probabilistic U processor
implementing all unitary transformations.

%%%%%%%%%%%%%%%%%%%%%%%%%%%%%%%%%%%%%%%%%%%%%%%%%%%%%%%%%%%%%%%
\section{Optimality for U(1) rotations}
%%%%%%%%%%%%%%%%%%%%%%%%%%%%%%%%%%%%%%%%%%%%%%%%%%%%%%%%%%%%%%%
In this section we will relax the universality condition and we will study an implementation
of a one-parametric group of unitary transformations. An example
of the universal processor performing this task is the CNOT processor. Our aim
will be to specify the dependence of the approximation
error and the success probability on the size $N$ of the program register.

Let us start with the approximative realization of
$U_\varphi=\exp(i\varphi A)$ on U processors, i.e.
we use processors of the form $G=\sum_j U_j\otimes |j\r\l j|$,
where $j=1,\dots,N$. Since the approximation
error is specified by overlaps between $U_\varphi$ and $U_j$, it is
reasonable to assume that $U_j$ are from the linear span of
unitaries $U_\varphi$. Otherwise, we would obtain
$\T U_\varphi^\dagger U_j=0$ which is not interesting for approximative
realization of $U_\varphi$, i.e. the associated states $|j\r$ do not
approximate any transformation $U_\varphi$.

In particular, $U_j$ are from the group $U_\varphi$. The overlap
between two unitaries $U_\varphi,U_{\varphi^\prime}$ is given
by the expression $|\T U_\varphi^\dagger
U_{\varphi^\prime}|^2=|\T \exp [i(\varphi^\prime-\varphi)A]|^2
=|\sum_n e^{i(\varphi^\prime-\varphi)a_n}|^2$, where $a_n$
are eigenvalues of $A$. The closer the angles $\varphi^\prime,\varphi$
are, the larger is the overlap. In what follows we will assume
that the data register is two-dimensional,
i.e. $A=\vec{a}\cdot\vec{\sigma}$,
$U_\varphi=\cos\varphi I+i\sin\varphi\,(\vec{a}\cdot\vec{\sigma})$
and $|\T U_\varphi^\dagger U_{\varphi^\prime}|^2=
4\cos^2(\varphi^\prime-\varphi)$.

Let us consider a processor with $N$ dimensional program register
performing $N$ unitaries $U_{\varphi_j}$. The question is,
what are the best choices of angles $\varphi_j$ in order
to minimize the approximation error of the implementation
of the whole group $U_\varphi$ ($\varphi\in[0,2\pi]$).
The program
state $|j\r$ best approximates those angles $\varphi$ that lie
in the vicinity of the angle $\varphi_j$ (the difference $\varphi_j-\varphi$ is smaller
than the differences $\varphi_k-\varphi$ for $k\ne j$).
Because of the
identity $\cos^2(\varphi_j-\varphi+\pi)=\cos^2(\varphi_j-\varphi)$
the state $|j\r$ approximates with the same accuracy the unitaries
$U_\varphi$ and $U_{\varphi+\pi}$. That is, it is enough to consider
only implementation of unitaries specified by angles within the
interval $\varphi\in[0,\pi]$. The whole problem can be illustrated
in the following way. The angles
$\varphi_j$ divide the half-circle (angles from 0 to $\pi$)
into $N$ regions. The state $|j\r$ approximates
optimally all the angles from the interval
$[\varphi_j-\frac{1}{2}(\varphi_{j}-\varphi_{j-1}),
\varphi_j+\frac{1}{2}(\varphi_{j+1}-\varphi_{j})]$
(for simplicity we assume that $\varphi_1=0$ and
$\varphi_j<\varphi_{j+1}$). The best choice of $\varphi_j$ is such that
the half-circle is divided into
equally large regions, i.e. the angles $\varphi_j$ are separated by the
same angle $\varphi_{j+1}-\varphi_j=\pi/N$. In such case the
approximation error reads
\be
\epsilon_G=1-\frac{1}{4}4\cos^2(\pi/(2N))=1-\cos^2(\pi/(2N))
\ee
because for arbitrary $\varphi$ the smallest difference is
$\min_j\{\varphi_j-\varphi\}\le\pi/(2N)$, i.e. the overlap is maximal
and the error $\epsilon_G$ is optimal. The average error of implementation
of the operations $U_\varphi$ equals to
\be
\overline{\epsilon}_G=\frac{1}{2}(1-\frac{N}{\pi}\sin\frac{\pi}{N}) \, .
\ee

Based on this result we can bound the optimal probabilistic realization
of  qubit operations $U_\varphi$ using the $N$ dimensional
program register and the U processor
\be
P_{\rm success}^G&\le& \cos^2(\frac{\pi}{2N}) \, , \\
\overline{P}_{\rm success}^G&\le& \frac{1}{2}(1+\frac{N}{\pi}\sin\frac{\pi}{N}) \, .
\ee
Now the question is whether there exists a measurement-assisted U processors
saturating these bounds. In what follows we will present an example
of the U processor that saturates this bound in a limit sense, i.e. for large
program registers.

In Refs. \cite{vidal1,vidal2} the authors proposed a way how to
increase the success probability to arbitrarily close to unity
by using large program registers with the dimensionality $N={\rm dim \cH_p}=2^n$, where
$n$ is the number of qubits in the program register.
They found a network of the controlled U form composed
of $n$ sequentially applied $k$-Toffoli gates for $k=1,2\dots n$.
The $k$-Toffoli ($T_k$) implements a controlled NOT operation
with $k$ control qubits and a single target qubit. The whole processor
acts on $n+1$ qubits, where the first qubit represents the data and all other
 qubits form the program register. The $k$-Toffoli gate
uses qubits $1,\dots k$ as control qubits
(this includes the program qubit as the qubit number 1)
and the $(k+1)$th qubit is the target qubit.
The whole processor is described by the operator
$G=T_n(T_{n-1}\otimes I)\dots (T_2\otimes I^{\otimes(n-2)} )
(T_1\otimes I^{\otimes(n-1)})$. Vidal et al. showed \cite{vidal1,vidal2}
that it is possible to choose program states and a measurement
such that the success probability scales as
\be
P_{\rm success}^M(U_\varphi)=1-(1/N)
\ee
for an arbitrary operation $U_\varphi=e^{i\varphi\sigma_z}$. Thus, we find that
$P_{\rm success}^{G,M}=\overline{P}_{\rm success}^{G,M}=1-(1/N)$, which
is consistent with the derived bound, but scales differently, because
$P_{success}^{G}\le 1-\sin^2(\pi/2N)\to 1-(\pi/(2N))^2$ for large $N$.
This means that the saturation of the bound is still an open question.
Moreover, let remind us that the derived formula for
the measurement-assisted U processor holds only for dimensions $N=2^n$,
where $n$ is the size of the program register in the number of qubits.

%%%%%%%%%%%%%%%%%%%%%%%%%%%%%%%%%%%%%%%%%%%%%%%%%%%%%%%%%%%%%
\section{Conclusion}
%%%%%%%%%%%%%%%%%%%%%%%%%%%%%%%%%%%%%%%%%%%%%%%%%%%%%%%%%%%%%
One of the goals of the research in the field of quantum computing
is a construction of programmable quantum processor.
It follows from the work of  Nielsen and Chuang
that perfect programmable processor  does not exist.
In this paper we have analyzed two different scenarios
relaxing the condition of universality: approximate
implementation of quantum programs (approximate processors)
and probabilistic implementation of quantum programs (probabilistic
processors).
We have discussed the problems of optimality
of universal approximate processors and universal
(measurement-assisted) probabilistic processors. Using
examples of the CNOT, the QID and the SWAP processors we
have shown explicitly the validity of the general relation between
the approximation error $\epsilon_G$ and the success probability
$P_{\rm success}^G$, respectively. In particular, we have
seen that the inequality $P_{\rm error}\ge\epsilon$ is not
saturated in general case. For instance, the SWAP processor
is completely useless in probabilistic programming
$P_{\rm error}^{\rm SWAP}=1$, but approximatively it performs optimally.

We have studied in detail optimality of restricted class of processors -
the so called U processors. We have shown that the QID processor is optimal
in implementing unitary transformations on a qudit with
$N=d^2$ dimensional program register. Under the same settings we have
analyzed the restricted universality of an implementation of only
one parametric set of unitary transformations. Unlike the QID
processor we have found that the presented example of the U processor
does not saturate the bound derived from approximative
implementation. That is, the question of optimality
for this processor is an open question.

The optimality questions of universal processors (either approximate, or
probabilistic) represent a difficult open problem of quantum information
science \cite{nielsen}.
In this paper, we have described  current state of the art
and we have presented known results. The question of
efficient programmability of quantum computers makes
these problems of optimality very attractive and they deserve
further investigation.

%%%%%%%%%%%%%%%%%%%%%%%%%%%%%%%%%%%%%%%%%%%%%%%%%%%%%%%%%%%%%%%%%%%%%%%%%%%%%

%%%%%%%%%%%%%%%%%%%%%%%%%%%%%%%%%%%%%%%%%%%%%%%% ACKNOWLDGMENT %%%%%%%%%%%%
\section*{Acknowledgements} This work was supported in part by the European
Union  projects INTAS-04-77-7289 and QAP,  by the Slovak
Academy of Sciences via the project CE-PI I/2/2005  and VEGA 2/6070/26, by the project
APVT-99-012304 and GA\v CR GA 201/01/0413.

%%%%%%%%%%%%%%%%%%%%%%%%%%%%%%%%%%%%%%%%%%%%%% BIBLIOGRAPHY %%%%%%%%%%%%%%%%

%%%%%%%%%%%%%%%%%%%%%%%%%%%%%%%%%%%%%%%%%%%%%%%%%%%%%%%%%%%%%%%%%%%%%%%%%%%%%


\begin{thebibliography}{00}

\bibitem{gruska}
J.\ Gruska, {\em Foundations of Computing} (Thompson Computer Press, London, 1999).
\bibitem{niel_chuang}
M.\ Nielsen and I.\ Chuang, Phys.\ Rev.\ Lett.\ {\bf 79}, 321 (1997).
\bibitem{hillery2002b}
M.\ Hillery, M.\ Ziman, and V.\ Bu\v{z}ek, Phys.\ Rev.\ A {\bf 66}, 042302 (2002). % deterministic processors 2002
\bibitem{vlasov}A.\ Yu.\ Vlasov, quant-ph/0103119; quant-ph/0311196; quant-ph/0301147; quant-ph/0503230.
\bibitem{dusek} M.\ Du\v{s}ek and V.\ Bu\v{z}ek, Phys.\ Rev.\ A {\bf 66}, 022112 (2002).
%\bibitem{fiurasek} J.\ Fiura\v{s}ek, M.\ Du\v{s}ek, and R.\ Filip, Phys.\ Rev.\ Lett.\ {\bf 89}, 190401 (2002).
\bibitem{dariano} G.\ M.\ D' Ariano and P.\ Perinotti, Phys.\ Rev.\ Lett. {\bf 94}, 090401 (2005).
\bibitem{geabanacloche1} J.\ Gea-Banacloche, Phys.\ Rev.\ A {bf 65}, 022308 (2002).
\bibitem{deutsch} A.\ Silberfarb and I.\ Deutsch, Phys.\ Rev.\ A {\bf 68}, 013817 (2003).
\bibitem{ekert} A.\ K.\ Ekert, C.\ M.\ Alves, D.\ K.\ L.\ Oi, M.\ Horodecki, P.\ Horodecki, and L.\ C.\ Kwek, Phys.\ Rev.\ Lett.\ {\bf 88}, 217901 (2002).
\bibitem{paz} J.\ P.\ Paz and A.\ Roncaglia, Phys.\ Rev.\ A {\bf 68}, 052316 (2003).
\bibitem{hillery2002a}
M.\ Hillery, M.\ Ziman, and V.\ Bu\v{z}ek, Phys.\ Rev.\ A {\bf 65}, 022301 (2002). % probabilistic processors 2002
\bibitem{hillery2004}
M.\ Hillery, M.\ Ziman, and V.\ Bu\v{z}ek, Phys.\ Rev.\ A {\bf 69}, 042311 (2004).
\bibitem{hillery2006}
M.\ Hillery, M.\ Ziman, and V.\ Bu\v{z}ek, Phys.\ Rev.\ A {\bf 73}, 022345 (2006).
\bibitem{dariano2005}
G.\ M.\ D' Ariano and P.\ Perinotti, quant-ph/0510033.
\bibitem{vidal1}
G.\ Vidal and J.\ I.\ Cirac, quant-ph/0012067.
\bibitem{vidal2}
G.\ Vidal, L. Masanes, and J.\ I.\ Cirac, Phys. Rev. Lett. {\bf 88}, 047905 (2002).
\bibitem{ziman2003}
M.\ Ziman and V.\ Bu\v zek, Int. J.Quant. Inf. {\bf 1}, 523 (2003)




\bibitem{nielsen}
M.A. Nielsen and I.L. Chuang,
{\it Quantum Computation and Quantum Information},
(Cambridge University Press, Cambridge, 2000)

\end{thebibliography}
\end{document}